# Detection of spin voltaic effect in a *p-n* heterojunction

T. Kondo*, J. Hayafuji, and H. Munekata

*Imaging Science and Engineering Laboratory, Tokyo Institute of Technology*

*4259-G2-13 Nagatsuta, Midori-ku Yokohama, 226-8502 Japan*

*Phone/Fax: +81-45-924-5179, e-mail: tkondo@isl.titech.ac.jp*

**Abstract**

Model calculation and experimental data of circularly-polarized-light-dependent photocurrent in a *n*-AlGaAs/*p*-InGaAs/*p*-GaAs heterostructure are reported. It is found that, under the appropriate forward bias condition, spin voltaic effect (SVE) can survive across the heterojunction and give rise to detectable polarization-dependent photocurrent signals which are greater than the signals due to the magnetic circular dichroism. Our analysis suggests that SVE can be enhanced by optimization of layer thickness, doping profile, and applied bias, making SVE favorable for the realization of a semiconductor-based polarization detector, a spin-photodiode (spin-PD).

* Corresponding Author: Tsuyoshi Kondo, Assistant Professor
Imaging Science and Engineering Laboratory,
Tokyo Institute of Technology
TEL/FAX; +81-45-924-5179;  e-mail; tkondo@isl.titech.ac.jp

Spintronics is a novel type of electronics that takes advantage of both charge and spin degrees of freedom. A spin-LED [1-4], which utilizes both spin-dependent optical and transport phenomena, is a semiconductor-based, prototype spin-optoelectronic device that converts an electrical signal into circularly polarized light. Eventually, a spin-LED is expected to find application as an electrical-switchable circularly polarized light emitter. Meanwhile, there were several reports on photo-induced electronic effects with circular polarized light [5-10] which could be used for the polarization detection. An optical system composed of a spin-LED and a polarization detector would be important for a future optical system, in that it could be used in unique applications in which circular dichroism provides valuable physical information. Example applications would include high-speed evaluations (MHz-GHz) of optical activity of molecules and nanostructures, magnetism in magnetic materials, and strains in solid structures; these are very important in the fields of pharmaceutics, electronics, and materials engineering. Polarization detection of a single photon [11] would be another important application.

This Letter is concerned with the detection of the spin voltaic effect (SVE) in a *p-n* heterojunction (a spin-photodiode). This effect was predicted by theoretically analyzing the diffusion-limited carrier transport in a *p-n* junction that has a spin-dependent diffusion potential [9,10]. Heterojunction with different *g*-factors (Fig.1(a)) is one of the realistic structures to pursue the SVE. Rigorously stated, however, it is not clear to what extent the abrupt band discontinuity at the heterointerface affects the spin-diffusion process. In addition, photocurrent originated from magnetic circular dichroism (MCD) should be distinguished from that due to SVE. With these two points in mind, we have examined theoretically the photocurrent characteristics of a *p-n* heterojunction under a magnetic field on the basis of established models of heterojunction transport [12] and a solar cell [13]. We have found that SVE can survive when forward bias becomes comparable or larger than the effective band discontinuity. We have also described the experimental detection of the SVE component in photocurrent in the *n*-(Al,Ga)As/*p*-(In,Ga)As/*p*-GaAs heterojunction. Our findings are very important in view of realization of diffusive spin transport devices.

Let us first consider diffusive spin/charge transport across a *p-n* heterojunction with a band edge profile shown in Fig.1(a). For simplicity, we restrict ourselves to electron transport and low-power excitation for which the number of background electrons in an *n*-type layer remains unchanged. When right-circularly-polarized ($\sigma^+$) light is illuminated on the *n*-layer [14], the spin polarization $P_n$ occurs in the *n*-layer, resulting in the spin-splitting of quasi Fermi levels ($\mu_\rightarrow$ and $\mu_\leftarrow$). At small forward bias, the number of spin-polarized electrons injected from *n*- to *p*-layer is substantially smaller than the number of electrons accumulated at the *p*-side of the heterointerface. In this case, the drop of the electron chemical potential $\delta\mu$ appears primarily at the heterointerface [12], and there is no spin-splitting of Fermi level in the *p*-layer. For large forward bias, the number of injected spin-polarized carriers increases exponentially and becomes high enough to affect the number of electrons in accumulation region. At this point, the diode starts to exhibit diffusion-limited transport characteristics [12]. Chemical potential gradient and the splitting of quasi Fermi levels both appear in *p*-layer, as drawn schematically in Fig.1(a). The onset of the diffusion-limited transport is given by $eV_f \sim \Delta E_c$ where $\Delta E_c = E_{c,n} - E_{c,p}$ at zero bias and $V_f$ being the applied forward bias. Consequently, for $eV_f > \Delta E_c$, an optically-induced spin polarization $P_n$ in the *n*-layer gives rise to an increase in the diffusive flow of right-spin electrons $F_{e\rightarrow}$ in a *p*-layer. Illumination with left-circularly-polarized ($\sigma^-$) light decreases $F_{e\rightarrow}$ and increases $F_{e\leftarrow}$, but $|\Delta F_{e\rightarrow}| > |\Delta F_{e\leftarrow}|$ so that current is decreased. As a whole, this diode outputs an electric current whose direction depends on the polarization of light.

A number of *n*-$Al_xGa_{1-x}As$/*p*-$In_yGa_{1-y}As$/*p*-GaAs heterojunctions was grown by molecular beam epitaxy on *p*-GaAs(001) substrates at a substrate temperature of $T_s = 520°C$, and processed into mesa-diodes by photolithography and standard wet chemical etching [15]. The Al and In contents were selected to be $x = 0.12$ and $y = 0.3$, respectively, for which $g \approx 0$ and $-1.9$ for the *n*- and the *p*-layers, respectively [16,17]. Thicknesses and doping concentrations were $d = 1$ μm, $N_D = 3 \times 10^{17}$ cm$^{-3}$ for the *n*-layer, and $d = 15$ nm, $N_A = 8 \times 10^{18}$ cm$^{-3}$ for the *p*-layer. $d$ and $N_D$ values in the *n*-layer were determined so as to suppress the influence of surface recombination and strong spin-relaxation in heavy *n*-type doping [18]. The band-edge discontinuities are $\Delta E_{c1} \sim 310$ meV for $Al_{0.12}Ga_{0.88}As/In_{0.3}Ga_{0.7}As$ and $\Delta E_{c2} \sim 220$ meV for $In_{0.3}Ga_{0.7}As$/GaAs. In view of Fig.1(a), we most likely have electron accumulation in an

In$_{0.3}$Ga$_{0.7}$As layer (a double-arrowhead symbol in the figure). The effective band discontinuity between *n*- and *p*-bulk regions at zero bias is inferred to be $\Delta E_c \sim 0.1$ eV.

Experiments were carried out at 4 K with magnetic fields applied perpendicular to the surface of the diodes which exhibited a rectifying *I-V* characteristic (Fig. 1(b)). Dependence of circularly polarization on photocurrent and photovoltage was measured by means of the lock-in technique combined with a stress modulated quarter wave plate (42 kHz). The wavelength and power of light illumination were $\lambda = 685$ nm and $P = 15$ mW/cm$^2$, respectively. We exclusively show in this Letter the difference in the photocurrent between σ$^+$ and σ$^-$ lights, $\Delta I = I_{\sigma+} - I_{\sigma-}$, as a function of magnetic field.

Polarization-dependent photocurrent due to SVE, $\Delta I_{SVE}$, is expressed as follows [9, 10]:

$$\Delta I_{SVE} = 2 \cdot \sinh(\frac{\Delta E(H)}{k_B \cdot T}) \cdot P_n \cdot I(V_f). \qquad (1)$$

Here, $\Delta E$ and $P_n$ are the magnitude of Zeeman splitting of the conduction band in a *p*-InGaAs layer and the spin polarization of electrons in a *n*-AlGaAs layer, respectively. $I(V_f)$ is a diffusive forward-current in the *p*-layer with an applied bias voltage $V_f$. The photocurrent due to magnetic-circular dichroism (MCD), $\Delta I_{MCD}$ can be expressed by modifying the standard solar cell model [13] by taking into account the difference in absorption coefficients between the σ$^+$ and σ$^-$ lights due to the Zeeman splitting at a given magnetic field; namely, $|\Delta\alpha(H)| = |\alpha_{\sigma+}(H) - \alpha_{H=0}| = |\alpha_{\sigma-}(H) - \alpha_{H=0}|$. In addition, the contribution from light and heavy hole bands should be weighted according to the selection rule of circularly polarized light absorption [13].

Consequently, the total $\Delta I_{MCD}$ is expressed as:

$$\Delta I_{MCD} = -2 \cdot \sum_{i=1}^{3} \frac{dI_{photo}(E_\lambda)}{d\alpha_i(E_\lambda)} \cdot \frac{B}{2 \cdot \sqrt{E_\lambda - E_{g,i}}} \cdot \left( \frac{1}{4}|g_{c,i}| - \frac{5}{4}|g_{v,i}| \right) \cdot \mu_B \cdot H \quad (2)$$

Here, $i$ = 1, 2, 3 represents a AlGaAs layer, a InGaAs layer, and a GaAs substrate, respectively. $E_\lambda$, $\mu_B$ and $H$ are photon energy of incident light, Bohr magneton and an applied magnetic field, respectively. $E_{g,i}$, $g_{c,i}$, $g_{v,i}$ are energy band-gap, electron and hole $g$-factors, respectively, of the $i$-th layer. $I_{photo}(E_\lambda)$ is the photocurrent ($H$ = 0) based on the depth profile of photogenerated carriers under the illumination with unpolarized light [13]. The term $dI_{photo}/d\alpha_i$ is the sensitivity of photocurrent with respect to MCD. Using the calculation protocol described in ref. 13, we calculated $I_{photo}(E_\lambda)$. For actual calculation, reduced absorption constant $B = 3 \times 10^4$ cm$^{-1}$ [19] and simplified $g$-value of $g_v$ = -2.8 [20] were used for all three constituent layers. Diffusion coefficients/lengths of electrons at low temperatures were pulled out from refs.21-23. Finally, the relation $\Delta I_{MCD}/I_{photo} = -7 \times 10^{-4}$ /Tesla was obtained for the present $p$-$n$ heterojunction diode. In this particular diode structure, the main $\Delta I_{MCD}$ component arises from the difference between the intensity of $\sigma^+$ and $\sigma^-$ light that reaches the $p$-GaAs substrate. The contribution of the $p$-InGaAs layer to $\Delta I_{MCD}$ is negligibly small because the layer is relatively thin.

Overall $I$-$V$ characteristics incorporating $\Delta I_{SVE}$ and $\Delta I_{MCD}$ are shown schematically in Fig. 2(a). The MCD effect appears itself as bias-independent photocurrent. In this particular case, $|\Delta I_{MCD}(\sigma^+)| > |\Delta I_{MCD}(\sigma^-)|$, as seen in the entire reverse bias region. The SVE component manifests itself in the photocurrent in the forward bias region. As discussed in the paragraph related to Fig.1(a), the $\sigma^+$-SVE photocurrent increases steeper than the $\sigma^-$-SVE photocurrent, giving rise to the crossover in *I-V* characteristics between the two opposite polarizations.

Crossover between the two effects can been seen much clearer in a $\Delta I$-$H$ curve. Figure 2(b) shows the results of simulation obtained for three different $P_n$ values at the forward bias of $V_f = 0.2$ V. At this bias condition, the tested diode exhibited a dark current of 43 nA and a photocurrent (unpolarized) of $I_{photo} = 0.2$ μA with a shunt resistance of 1 MΩ. For $P_n = 5 \times 10^{-4}$, $\Delta I$ decreases monotonously with increasing $H$, indicating that $\Delta I_{MCD}$ is the dominant component. For $P_n = 1.5 \times 10^{-3}$, $\Delta I_{SVE}$ becomes more influential than $\Delta I_{MCD}$ under high magnetic fields, exhibiting a reversal of the slope in the $\Delta I$-$H$ curve at around 2 T. For $P_n = 3 \times 10^{-3}$, $\Delta I$ is dominated by the non-linear $\Delta I_{SVE}$ component. For $P_n = 0.5$, $\Delta I_{SVE}$ is 100 times larger than that for $P_n = 3 \times 10^{-3}$ (not shown in the figure).

Figure 3 shows experimental data obtained from the tested diode with a shunt resistance of 1 MΩ. The slope of the $\Delta I - H$ curve is negative within the range of $H \leq \pm 2$ T, and turns positive with further increasing $H$, indicating that the tested diode indeed outputs a current due to the SVE at relatively high magnetic field region. The fit to the experimental curve using eq.(1) yields $P_n = 1.4 \times 10^{-3}$ assuming that the dark current $I = 43$ nA consists of the diffusion current. Because of small signals, one may wonder that the current component which is not relevant to the diffusive transport is inadvertently misinterpreted in terms of SVE. We point out, however, that the contribution of such current components, *e.g.* recombination and/or parasitic leak current, would only show even-functional behavior with respect to magnetic fields. Tunneling current through the triangular barrier at the AlGaAs/InGaAs interface hardly exists because of a thick barrier width due to the low-doping level of the *n*-layer.

In summary, model calculation and experimental data of circularly-polarized-light-dependent photocurrent in a *n*-AlGaAs/*p*-InGaAs/*p*-GaAs heterostructure are described. The results indicate that, under the appropriate forward bias condition, spin voltaic effect (SVE) can survive across the heterojunction and give rise to detectable polarization-dependent photocurrent signals which are greater than the signals due to the magnetic circular dichroism. Our analysis suggests that SVE can be enhanced by optimization of layer thickness, doping profile, and applied bias, all of which are familiar parameters in semiconductor technology. Another point to enhance SVE is to utilize materials with larger spin splitting; e.g., narrow-gap III-Sb [17] and ferromagnetic/paramagnetic semiconductors.

The authors are grateful to Dr. I. Zutic of State University of New York at Buffalo for various discussions. This work is supported in part by Grant-in-Aid for Scientific Research from MEXT and JSPS (No. 14076210, No. 17760008, and No. 17206002), and by the NSF-IT program (DMR-0325474) in collaboration with Rice University.

**References**

[1]  Y. Ohno, D. K. Young, B. Beschoten, F. Matsukura, H. Ohno, D. D. Awschalom, Nature **402**, 790 (1999).

[2]  R. Fiederling, M. Keim, G. Reuscher, W. Ossau, G. Schmidt, A. Waag, L. W. Molenkamp, Nature **402**, 787 (1999).

[3]  B. T. Jonker, Proceedings of the IEEE **91**, 727 (2003).

[4]  X. Jiang, R. Wang, R. M. Shelby, R. M. Macfarlane, S. R. Bank, J. S. Harris, and S. S. P. Parkin, Phys. Rev. Lett. **94**, 056601 (2005).

[5]  A. F. Isakovic, D. M. Carr, J. Strand, B. D. Schultz, C. J. Palmstrom, and P. A. Crowell, J. Appl. Phys. **91**, 7261 (2002).

[6]  T. Manago, Y. Suzuki, and E. Tamura, J. Appl. Phys. **91**, 10130 (2002).

[7]  T. Taniyama, G. Wastlbauer, A. Ionescu, M. Tselepoi, and J. A. C. Bland, Phys. Rev. B **68**, 134430 (2003).

[8]  S. D. Ganichev, E. L. Ivchenko, V. V. Bel'kov, S. A. Tarasenko, M. Sollinger, D. Weiss, W. Wegscheider, and W. Prettl, Nature **417**, 153 (2002).

[9]  I. Zutic, J. Fabian, and S. Das Sarma, Phys. Rev. Lett. **88**, 066603 (2002).

[10]  J. Fabian, I. Zutic, and S. Das Sarma, Phys. Rev. B **66**, 165301 (2002).

[11]  I. Walmsley and P. Knight, Optics & Photonics News **13**, 43 (2002).

[12]  A. A. Grinberg and S. Luryi, IEEE Trans. Electron Device **40**, 859 (1993).

[13] S. M. Sze, *Physics of Semiconductor devices 2nd edition,* (John Wiley & Sons, New York, 1981), chap. 14.

[14] M. I. Dyakonov and V. I. Perel, in *Optical Orientation*, edited by F. Meier, B. P. Zakharchenya (North-Holland, Amsterdam, 1984), Chap. 1.

[15] J. Hayafuji, T. Kondo, and H. Munekata, IOP Conference Proceeding Series **184**, 127. (2005).

[16] C. Weisbuch and C. Hermann, Phys. Rev. B **15**, 816. (1977).

[17] H. Kosaka, A. A. Kiselev, F. A. Baron, K. W. Kim, E. Yablonovitch, Electron Lett. **37**, 464 (2001).

[18] R. I. Dzhioev, K. V. Kavokin, V. L. Korenev, M. V. Lazarev, B. Ya. Meltser, M. N. Stepanova, B. P. Zakharchenya, D. Gammon, and D. S. Katzer, Phys. Rev. B **66**, 245204 (2002).

[19] R. A. Smith, *Semiconductor 2nd edition,* (Cambridge University Press, London, 1978), chap. 10.

[20] Ch. Neumann, A. Nothe, and N. O. Lipari, Phys. Rev. B **37**, 922 (1988).

[21] E. H. Stevens, S. S. Yee, J. Appl. Phys. **44**, 715 (1973).

[22] W. Walukiewicz, J. Lagowski, L. Jastrzebski, and H. C. Gatos, J. Appl. Phys. **50**, 5040 (1979).

[23] D. J. Wolford, G. D. Gilliland, T. F. Kuech, L. M. Smith, J. Martinsen, J. A. Bradley, C. F.

Tsang, R. Venkatasubramanian, S. K. Ghandi, H. P. Hjaimarson, J. Vac. Sci. Technol. B **9**, 2369 (1991).

**Figure captions**

Fig. 1: (a) Schematic illustration of band edge profiles and quasi Fermi levels across the *p-n* heterojunction with different Zeeman splitting. Right-circularly-polarized light $\sigma^+$ is illuminated from the left, whereas the direction of an external field *H* is antiparallel to the propagation direction of light. A large enough forward bias $V_f$ is applied, showing the diffusion limited spin/carrier transport with electron flow $F_e$. $n_n$, $n_p$, $n_{p0}$, $\mu$, $\delta\mu$, are electron distributions during the illumination in *n*- and *p*-layers, electron distribution in equilibrium in a *p*-layer, quasi Fermi levels, and a discontinuity of Fermi level at the heterointerface. Small arrows in super-subscript represent direction of electron spin. $E_{c,p}$ and $E_{n,p}$ are the band edge of *n*- and *p*-layers. A double-arrowhead symbol represents the region of electron accumulation. (b) An *I-V* curve of the tested *n*-AlGaAs/*p*-InGaAs/*p*-GaAs diode in the dark at 4 K. The diameter of a mesa-diode is 300 μm.

Fig.2: (a) Schematic *I-V* characteristics showing relationship between $\Delta I$, $\Delta I_{MCD}$, and $\Delta I_{SVE}$, under the illumination with circularly-polarized light $\sigma^+$ and $\sigma^-$, and (b) calculated $\Delta I - H$ curves with three different $P_n$ values. The device parameters ($D_e$, $D_h$, $L_e$, and $L_h$) used for the estimation of $\Delta I_{MCD}$ are $D_h = 0.035$ cm$^2$/s, $L_h = 180$ nm for the *n*-layer, and $D_e = 0.70$ cm$^2$/s, $L_e = 830$ nm for the *p*-layer.

Fig.3: $\Delta I$-$H$ characteristics of a tested $n$-$Al_{0.12}Ga_{0.88}As$/$p$-$In_{0.3}Ga_{0.7}As$/$p$-GaAs diode at 4 K. Diameter of a top optical-access window of a diode is 240 μm. A solid line represents a calculated curve with $P_n = 1.4 \times 10^{-3}$.

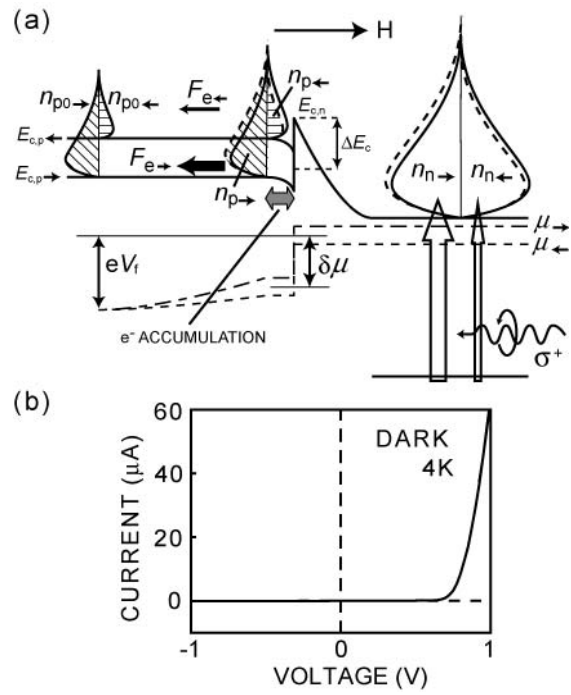

Fig. 1

T. Kondo *et al.*

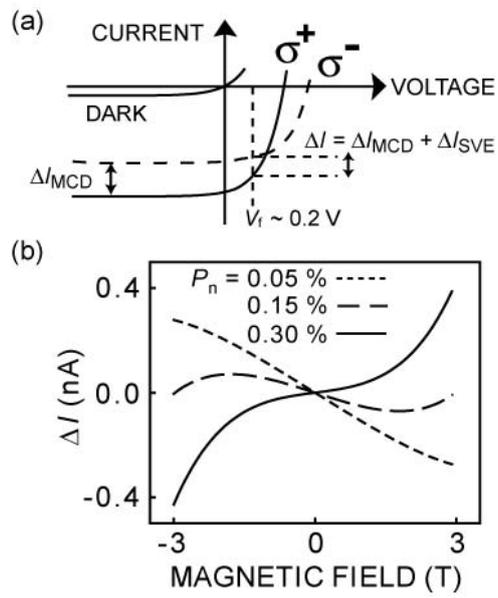

Fig. 2

T. Kondo et al.

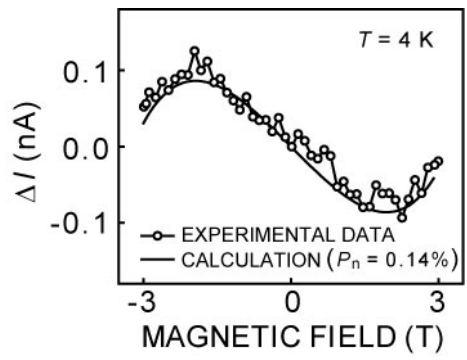

Fig. 3

T. Kondo *et al.*